\RequirePackage{lineno}

\documentclass[aps,prl,twocolumn,showpacs,superscriptaddress,letterpaper,amsmath,amssymb]{revtex4}

\usepackage{graphicx}  % needed for figures
\usepackage{dcolumn}   % needed for some tables
\usepackage{bm}        % for math
\usepackage{amssymb}   % for math
\usepackage{feynmf}%%%{feynmp}
\usepackage{slashed}
\def\gsim{\lower0.5ex\hbox{$\:\buildrel >\over\sim\:$}}
\def\lsim{\lower0.5ex\hbox{$\:\buildrel <\over\sim\:$}}

\def \n{\noindent}

\let\n=\nu
\let\t=\tau

\newcommand{\be}{\begin{equation}}
\newcommand{\ee}{\end{equation}}
\newcommand{\bea}{\begin{eqnarray}}
\newcommand{\eea}{\end{eqnarray}}

\newcommand{\nbox}{{\,\lower0.9pt\vbox{\hrule \hbox{\vrule height 0.2 cm
\hskip 0.2 cm \vrule height 0.2 cm}\hrule}\,}}

\begin{document}
%\pagewiselinenumbers

\thispagestyle{empty}
\vspace*{-3.5cm}

%\begin{flushright}
%\today\\
%\end{flushright}
%\vspace{0.5in}
%\renewcommand{\baselinestretch}{2}

\bibliographystyle{apsrev}

\title{{Search for pair-production of strongly-interacting particles \\ decaying to
    pairs of jets in {\bf $p\bar{p}$} collisions at {\bf $\sqrt{s}=1.96$ TeV}}}

\affiliation{Institute of Physics, Academia Sinica, Taipei, Taiwan 11529, Republic of China}
\affiliation{Argonne National Laboratory, Argonne, Illinois 60439, USA}
\affiliation{University of Athens, 157 71 Athens, Greece}
\affiliation{Institut de Fisica d'Altes Energies, ICREA, Universitat Autonoma de Barcelona, E-08193, Bellaterra (Barcelona), Spain}
\affiliation{Baylor University, Waco, Texas 76798, USA}
\affiliation{Istituto Nazionale di Fisica Nucleare Bologna, $^{ee}$University of Bologna, I-40127 Bologna, Italy}
\affiliation{University of California, Davis, Davis, California 95616, USA}
\affiliation{University of California, Los Angeles, Los Angeles, California 90024, USA}
\affiliation{Instituto de Fisica de Cantabria, CSIC-University of Cantabria, 39005 Santander, Spain}
\affiliation{Carnegie Mellon University, Pittsburgh, Pennsylvania 15213, USA}
\affiliation{Enrico Fermi Institute, University of Chicago, Chicago, Illinois 60637, USA}
\affiliation{Comenius University, 842 48 Bratislava, Slovakia; Institute of Experimental Physics, 040 01 Kosice, Slovakia}
\affiliation{Joint Institute for Nuclear Research, RU-141980 Dubna, Russia}
\affiliation{Duke University, Durham, North Carolina 27708, USA}
\affiliation{Fermi National Accelerator Laboratory, Batavia, Illinois 60510, USA}
\affiliation{University of Florida, Gainesville, Florida 32611, USA}
\affiliation{Laboratori Nazionali di Frascati, Istituto Nazionale di Fisica Nucleare, I-00044 Frascati, Italy}
\affiliation{University of Geneva, CH-1211 Geneva 4, Switzerland}
\affiliation{Glasgow University, Glasgow G12 8QQ, United Kingdom}
\affiliation{Harvard University, Cambridge, Massachusetts 02138, USA}
\affiliation{Division of High Energy Physics, Department of Physics, University of Helsinki and Helsinki Institute of Physics, FIN-00014, Helsinki, Finland}
\affiliation{University of Illinois, Urbana, Illinois 61801, USA}
\affiliation{The Johns Hopkins University, Baltimore, Maryland 21218, USA}
\affiliation{Institut f\"{u}r Experimentelle Kernphysik, Karlsruhe Institute of Technology, D-76131 Karlsruhe, Germany}
\affiliation{Center for High Energy Physics: Kyungpook National University, Daegu 702-701, Korea; Seoul National University, Seoul 151-742, Korea; Sungkyunkwan University, Suwon 440-746, Korea; Korea Institute of Science and Technology Information, Daejeon 305-806, Korea; Chonnam National University, Gwangju 500-757, Korea; Chonbuk National University, Jeonju 561-756, Korea; Ewha Womans University, Seoul, 120-750, Korea}
\affiliation{Ernest Orlando Lawrence Berkeley National Laboratory, Berkeley, California 94720, USA}
\affiliation{University of Liverpool, Liverpool L69 7ZE, United Kingdom}
\affiliation{University College London, London WC1E 6BT, United Kingdom}
\affiliation{Centro de Investigaciones Energeticas Medioambientales y Tecnologicas, E-28040 Madrid, Spain}
\affiliation{Massachusetts Institute of Technology, Cambridge, Massachusetts 02139, USA}
\affiliation{Institute of Particle Physics: McGill University, Montr\'{e}al, Qu\'{e}bec H3A~2T8, Canada; Simon Fraser University, Burnaby, British Columbia V5A~1S6, Canada; University of Toronto, Toronto, Ontario M5S~1A7, Canada; and TRIUMF, Vancouver, British Columbia V6T~2A3, Canada}
\affiliation{University of Michigan, Ann Arbor, Michigan 48109, USA}
\affiliation{Michigan State University, East Lansing, Michigan 48824, USA}
\affiliation{Institution for Theoretical and Experimental Physics, ITEP, Moscow 117259, Russia}
\affiliation{University of New Mexico, Albuquerque, New Mexico 87131, USA}
\affiliation{The Ohio State University, Columbus, Ohio 43210, USA}
\affiliation{Okayama University, Okayama 700-8530, Japan}
\affiliation{Osaka City University, Osaka 588, Japan}
\affiliation{University of Oxford, Oxford OX1 3RH, United Kingdom}
\affiliation{Istituto Nazionale di Fisica Nucleare, Sezione di Padova-Trento, $^{ff}$University of Padova, I-35131 Padova, Italy}
\affiliation{University of Pennsylvania, Philadelphia, Pennsylvania 19104, USA}
\affiliation{Istituto Nazionale di Fisica Nucleare Pisa, $^{gg}$University of Pisa, $^{hh}$University of Siena and $^{ii}$Scuola Normale Superiore, I-56127 Pisa, Italy, $^{mm}$INFN Pavia and University of Pavia, I-27100 Pavia, Italy}
\affiliation{University of Pittsburgh, Pittsburgh, Pennsylvania 15260, USA}
\affiliation{Purdue University, West Lafayette, Indiana 47907, USA}
\affiliation{University of Rochester, Rochester, New York 14627, USA}
\affiliation{The Rockefeller University, New York, New York 10065, USA}
\affiliation{Istituto Nazionale di Fisica Nucleare, Sezione di Roma 1, $^{jj}$Sapienza Universit\`{a} di Roma, I-00185 Roma, Italy}
\affiliation{Texas A\&M University, College Station, Texas 77843, USA}
\affiliation{Istituto Nazionale di Fisica Nucleare Trieste/Udine; $^{nn}$University of Trieste, I-34127 Trieste, Italy; $^{kk}$University of Udine, I-33100 Udine, Italy}
\affiliation{University of Tsukuba, Tsukuba, Ibaraki 305, Japan}
\affiliation{Tufts University, Medford, Massachusetts 02155, USA}
\affiliation{University of Virginia, Charlottesville, Virginia 22906, USA}
\affiliation{Waseda University, Tokyo 169, Japan}
\affiliation{Wayne State University, Detroit, Michigan 48201, USA}
\affiliation{University of Wisconsin, Madison, Wisconsin 53706, USA}
\affiliation{Yale University, New Haven, Connecticut 06520, USA}

\author{T.~Aaltonen}
\affiliation{Division of High Energy Physics, Department of Physics, University of Helsinki and Helsinki Institute of Physics, FIN-00014, Helsinki, Finland}
\author{E.~Albin}
\affiliation{University of California Irvine, Irvine, CA 92697, USA}
\author{S.~Amerio}
\affiliation{Istituto Nazionale di Fisica Nucleare, Sezione di Padova-Trento, $^{ff}$University of Padova, I-35131 Padova, Italy}
\author{D.~Amidei}
\affiliation{University of Michigan, Ann Arbor, Michigan 48109, USA}
\author{A.~Anastassov$^x$}
\affiliation{Fermi National Accelerator Laboratory, Batavia, Illinois 60510, USA}
\author{A.~Annovi}
\affiliation{Laboratori Nazionali di Frascati, Istituto Nazionale di Fisica Nucleare, I-00044 Frascati, Italy}
\author{J.~Antos}
\affiliation{Comenius University, 842 48 Bratislava, Slovakia; Institute of Experimental Physics, 040 01 Kosice, Slovakia}
\author{G.~Apollinari}
\affiliation{Fermi National Accelerator Laboratory, Batavia, Illinois 60510, USA}
\author{J.A.~Appel}
\affiliation{Fermi National Accelerator Laboratory, Batavia, Illinois 60510, USA}
\author{T.~Arisawa}
\affiliation{Waseda University, Tokyo 169, Japan}
% ...

\collaboration{CDF Collaboration\footnote{With visitors from
$^a$Istituto Nazionale di Fisica Nucleare, Sezione di Cagliari, 09042 Monserrato (Cagliari), Italy,
$^b$University of California Irvine, Irvine, CA 92697, USA,
$^e$Institute of Physics, Academy of Sciences of the Czech Republic, 182~21, Czech Republic,
$^f$CERN, CH-1211 Geneva, Switzerland,
$^g$Cornell University, Ithaca, NY 14853, USA,
$^h$University of Cyprus, Nicosia CY-1678, Cyprus,
$^i$Office of Science, U.S. Department of Energy, Washington, DC 20585, USA,
$^j$University College Dublin, Dublin 4, Ireland,
$^k$ETH, 8092 Z\"{u}rich, Switzerland,
$^l$University of Fukui, Fukui City, Fukui Prefecture, Japan 910-0017,
$^m$Universidad Iberoamericana, Lomas de Santa Fe, M\'{e}xico, C.P. 01219, Distrito Federal,
$^n$University of Iowa, Iowa City, IA 52242, USA,
$^o$Kinki University, Higashi-Osaka City, Japan 577-8502,
$^p$Kansas State University, Manhattan, KS 66506, USA,
$^q$Brookhaven National Laboratory, Upton, NY 11973, USA,
$^r$University of Manchester, Manchester M13 9PL, United Kingdom,
$^s$Queen Mary, University of London, London, E1 4NS, United Kingdom,
$^t$University of Melbourne, Victoria 3010, Australia,
$^u$Muons, Inc., Batavia, IL 60510, USA,
$^v$Nagasaki Institute of Applied Science, Nagasaki 851-0193, Japan,
$^w$National Research Nuclear University, Moscow 115409, Russia,
$^x$Northwestern University, Evanston, IL 60208, USA,
$^y$University of Notre Dame, Notre Dame, IN 46556, USA,
$^z$Universidad de Oviedo, E-33007 Oviedo, Spain,
$^{aa}$CNRS-IN2P3, Paris, F-75205 France,
$^{cc}$Universidad Tecnica Federico Santa Maria, 110v Valparaiso, Chile,
$^{dd}$Yarmouk University, Irbid 211-63, Jordan,
$^{ll}$Universite catholique de Louvain, 1348 Louvain-La-Neuve, Belgium,
$^{oo}$University of Z\"{u}rich, 8006 Z\"{u}rich, Switzerland,
$^{pp}$Massachusetts General Hospital and Harvard Medical School, Boston, MA 02114 USA,
$^{qq}$Hampton University, Hampton, VA 23668, USA,
$^{rr}$Los Alamos National Laboratory, Los Alamos, NM 87544, USA
}}
\noaffiliation

\begin{abstract}
We present a search for the pair-production of a non-standard-model strongly-interacting
particle that decays
 to a pair of quarks or gluons, leading to a final state with four
 hadronic jets. We consider both non-resonant production via an intermediate gluon
 as well as resonant production  via a distinct non-standard-model intermediate strongly-interacting
 particle.  We use
 data collected by the CDF experiment in proton-antiproton collisions at
 $\sqrt{s}=1.96$ TeV corresponding to an integrated luminosity of 6.6
 fb$^{-1}$. We find the data to be consistent with 
 standard model predictions. We report limits on $\sigma(p\bar{p}\rightarrow$
 {\it jjjj}$)$ as a function of the masses of the hypothetical
 intermediate particles. Upper limits  on the production cross
 sections for
 non-standard-model particles in several resonant and non-resonant
 processes are also derived.
% In the
% context of non-resonant
% coloron $C$ production, we exclude $m_C$ between 50 and 125
% GeV$/c^2$; in an $R$-parity-violating stop quark $\tilde{t}$ hypothesis we
% exclude $m_{\tilde{t}}$ between 50 and 100 GeV$/c^2$. In the context
% of hyperpions produced via an axi-gluon  resonance $A$, we exclude
% $m_A$ between 100 and 400 GeV$/c^2$ for coupling values preferred by
% the anomaly in the CDF top-quark forward-backward asymmetry.
\end{abstract}

% 12.60.-i Models beyond the standard model
% 13.85.Rm Limits on production of particles
% 14.65.-q Quarks
% 14.80.-j Other particles (including hypothetical)


\maketitle

\date{\today}

One of the few hints of possible physics beyond the standard model (SM)  at the TeV
scale is the anomalous 
 top-quark forward-backward asymmetry $A_{\it{fb}}$
  observed at the Tevatron~\cite{afb1,afb2,afb3}. 
% Many theoretical
%models have been proposed to explain this result, most with specific
% predictions of excesses which would confirm their
% hypotheses. However, nearly all have been excluded, mostly due to the lack of
%excesses of same-sign top-quark pair production~\cite{sstopatlas,sstopcms}, or top+jet
%resonances~\cite{tjcdf,tjcms,tjatlas}. 
This asymmetry could be generated by non-SM physics through the production of
top-quark pairs via a light axi-gluon~\cite{axidef}, a particle with axial couplings
to quarks, that interferes with standard
model (SM) $t\bar{t}$
production to produce the observed asymmetry. The axi-gluon would be
visible in its alternate decay mode to low-mass strongly-interacting particles, each
of which decays to a pair of jets~\cite{axigluon1} yielding a four-jet
final state.  This final state is of broad interest, as various models
predict pair-production of strongly-interacting particles decaying to jet pairs
with no intermediate resonance~\cite{dobrescu,kilic} and  $R$-parity-violating supersymmetric theories~\cite{susy}
predict pair-production of light partners of the top quark (stop
quarks),  each decaying into to pairs of light quarks.

The masses of the axi-gluon and its strongly-interacting decay products are not
predicted, but must be fairly light ($< 400$ GeV/$c^2$) to explain the $A_{\it{fb}}$
measurement~\cite{axigluon2}.  The LHC experiments have excellent sensitivity at high
mass due to the large center-of-mass energy, but difficulties at low
mass due to high background rates. The ATLAS experiment ruled out masses
between 100 and 150 GeV$/c^2$~\cite{atlas}; CMS ruled out masses
between 250 and 740 GeV$/c^2$~\cite{cms}.  No experimental bounds
exist for such non-SM particles with masses below 100
GeV$/c^2$ for non-resonant pair-production of di-jet resonances; there
are no current limits on resonant production.

\begin{figure}
\includegraphics[width=1.4in]{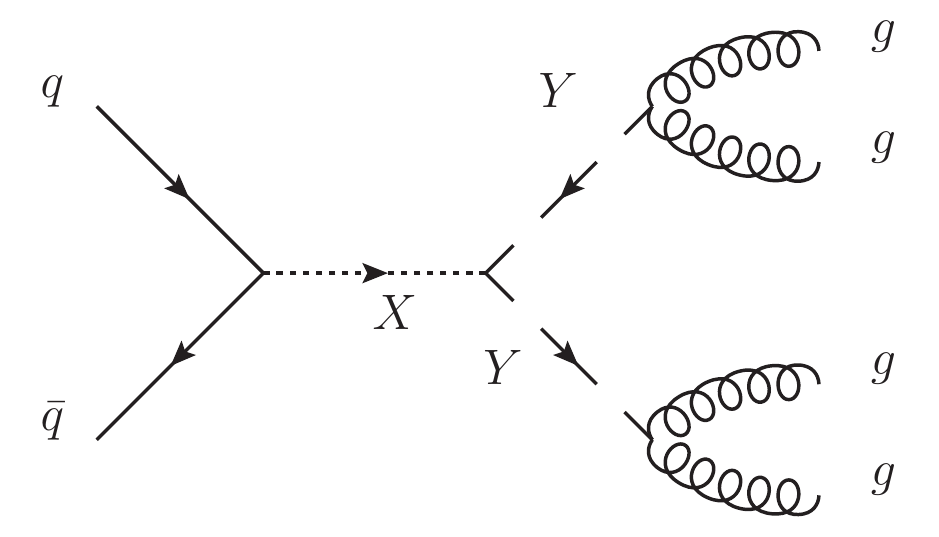}
\includegraphics[width=1.4in]{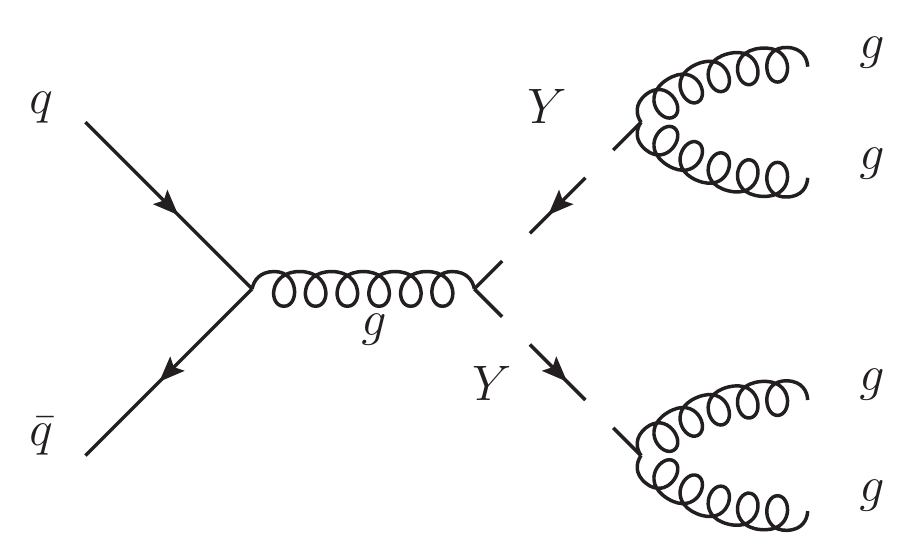}\ \ \ \ \ 
\caption{Diagrams for resonant (left, via $X$) and non-resonant
  (right) pair-production of $Y$ particles, with subsequent decays to pairs of
  gluons. Other models, with final-state quarks, are also considered.}
\label{fig:diag}
\end{figure}

In this Letter we report a search for both non-resonant and resonant
production of pairs of strongly-interacting particles, each of which decays to a pair of jets.
Rather than probing a specific theory, we construct a simplified model
with the minimal particle content. In the non-resonant case, we
consider the  production process $p\bar{p}\rightarrow YY \rightarrow
${\it jj jj}, with
 the mass of the hypothetical $Y$ state, $m_Y$ as a single free
 parameter.  In the resonant case, $p\bar{p}\rightarrow X
\rightarrow YY \rightarrow${\it jj jj}, we also explore the mass of the
$X$ state, $m_X$ (Fig.~\ref{fig:diag}). In both cases, we assume that the
natural width of the particles is small compared to the experimental resolution.

 We analyze a sample of events corresponding to an integrated
luminosity of 6.6$\pm 0.5$ fb$^{-1}$ recorded by the CDF II
detector~\cite{cdf}, a general purpose detector designed to study
$p\overline{p}$ collisions at $\sqrt{s}=1.96$ TeV produced by the Fermilab Tevatron collider. 
The tracking system consists of a silicon microstrip tracker and a
drift chamber  immersed in a 1.4 T axial magnetic field~\cite{trackingsystems}. 
Electromagnetic and hadronic calorimeters surrounding
the tracking system measure particle energies, with muon detection
provided by an additional system of drift chambers
located outside the calorimeters.

We reconstruct jets in the calorimeter using the {\sc
  jetclu}~\cite{jetclu} algorithm with a clustering radius of 0.4 in $\eta-\phi$ space~\cite{coordinates}, and calibrated using
the techniques outlined in Ref.~\cite{jes}.   Events are selected
online (triggered) by the requirement of three jets, each with
$E_{T}>20$ GeV and with $\Sigma_{\textrm{jets}} E_T>130$
GeV~\cite{coordinates}. The data set used in this search is limited to
6.6 fb$^{-1}$ because the trigger selection was not available in early
data. After trigger selection,
events are retained if at least four jets are found with $E_T>15$ GeV and
$|\eta|<2.4$.

We model  resonant and non-resonant production  with
{\sc madgraph5}~\cite{madgraph} version 1.4.8.4 and the {\sc cteq6l1}~\cite{cteq} parton
distribution functions (PDF). Additional parton radiation, hadronization, and
underlying-event modeling are described by {\sc pythia}~\cite{pythia} version 6.420. The detector
response for all simulated samples is modeled by the {\sc geant}-based CDF II detector simulation~\cite{cdfsimul}.

\begin{figure}
\includegraphics[width=0.9\linewidth]{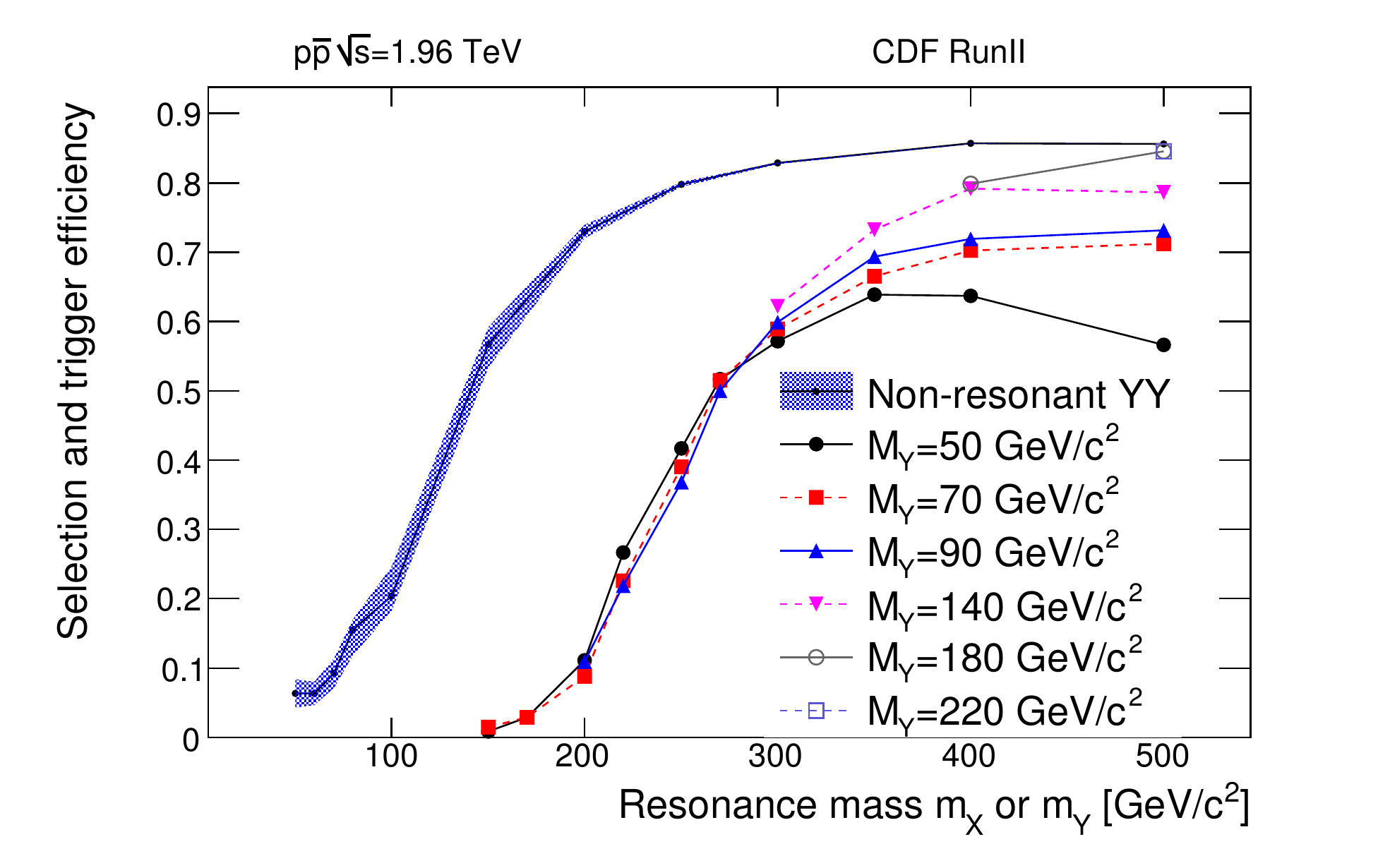}
\caption{ Overall efficiency, including trigger and selection
  requirements. Efficiency is shown for several simulated
  non-resonant $YY\rightarrow${\it jj jj} samples with varying
  $m_Y$. The shaded band shows the uncertainty. In addition,
  efficiency is shown for several simulated
  resonant $X\rightarrow YY\rightarrow${\it jj jj} samples with varying
  $m_X$ and $m_Y$. The uncertainty is not shown but is similar to the
  non-resonant case. The turn-on curve is determined largely by the
  trigger requirement that $\Sigma_{\textrm{jets}} E_T>130$ GeV. }
\label{fig:eff}
\end{figure}

The trigger and
selection requirements have an efficiency on the signal up to 90\% if
$\Sigma_{\textrm{jets}} E_T$ exceeds significantly the $130$ GeV trigger threshold. For events with smaller
$\Sigma_{\textrm{jets}} E_T$, the efficiency decreases rapidly (Fig.~\ref{fig:eff}). In the non-resonant-production model, the  $\Sigma_{\textrm{jets}} E_T$ is
strongly correlated with $m_Y$.  In the resonant-production model it is
correlated with $m_X$; additionally if $m_X-2m_Y$ is large, the $p_T$
of the resulting $Y$ is large, which leads to a small opening angle of its decay
products and a loss of efficiency due to merged jets. The trigger
efficiency is measured in simulated events, and uncertainties derived
from validation in disjoint samples; the measured
trigger efficiency and uncertainty are applied to the signal model.

% resonance reconstruction

\begin{figure}
\begin{picture}(90,60)
\put(0,0){\includegraphics[width=1.1\linewidth]{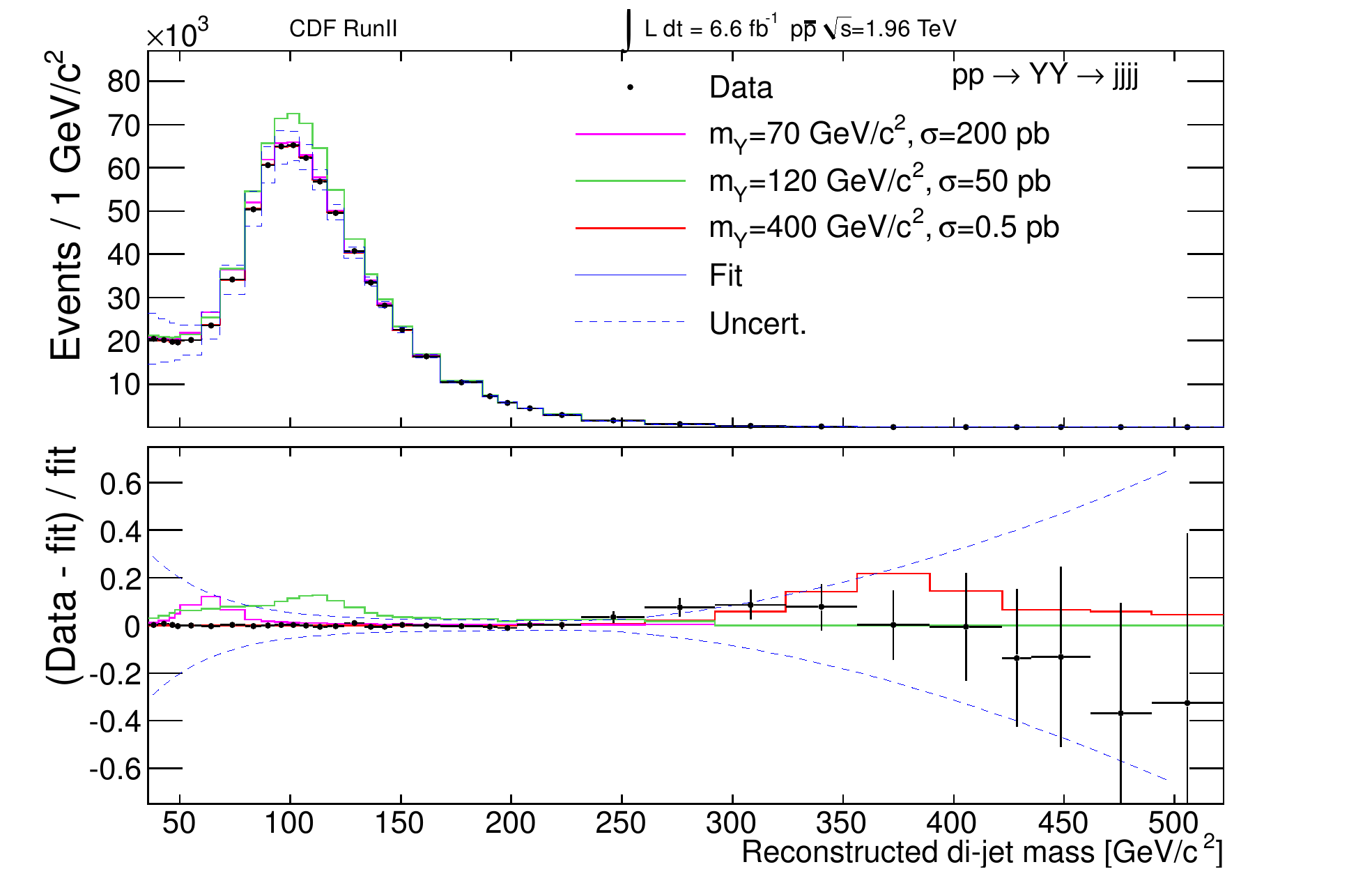}}
\put(12,52){(a)}
\put(12,24){(b)}
\end{picture}
\caption{Reconstructed mean di-jet mass in events with four jets. Parametric fit and several signal
  hypotheses overlaid in (a). Relative difference
  between the observed data and the fit in each bin shown in (b).}
\label{fig:mass2j}
\end{figure}

\begin{figure}
\begin{picture}(90,60)
\put(0,0){\includegraphics[width=1.1\linewidth]{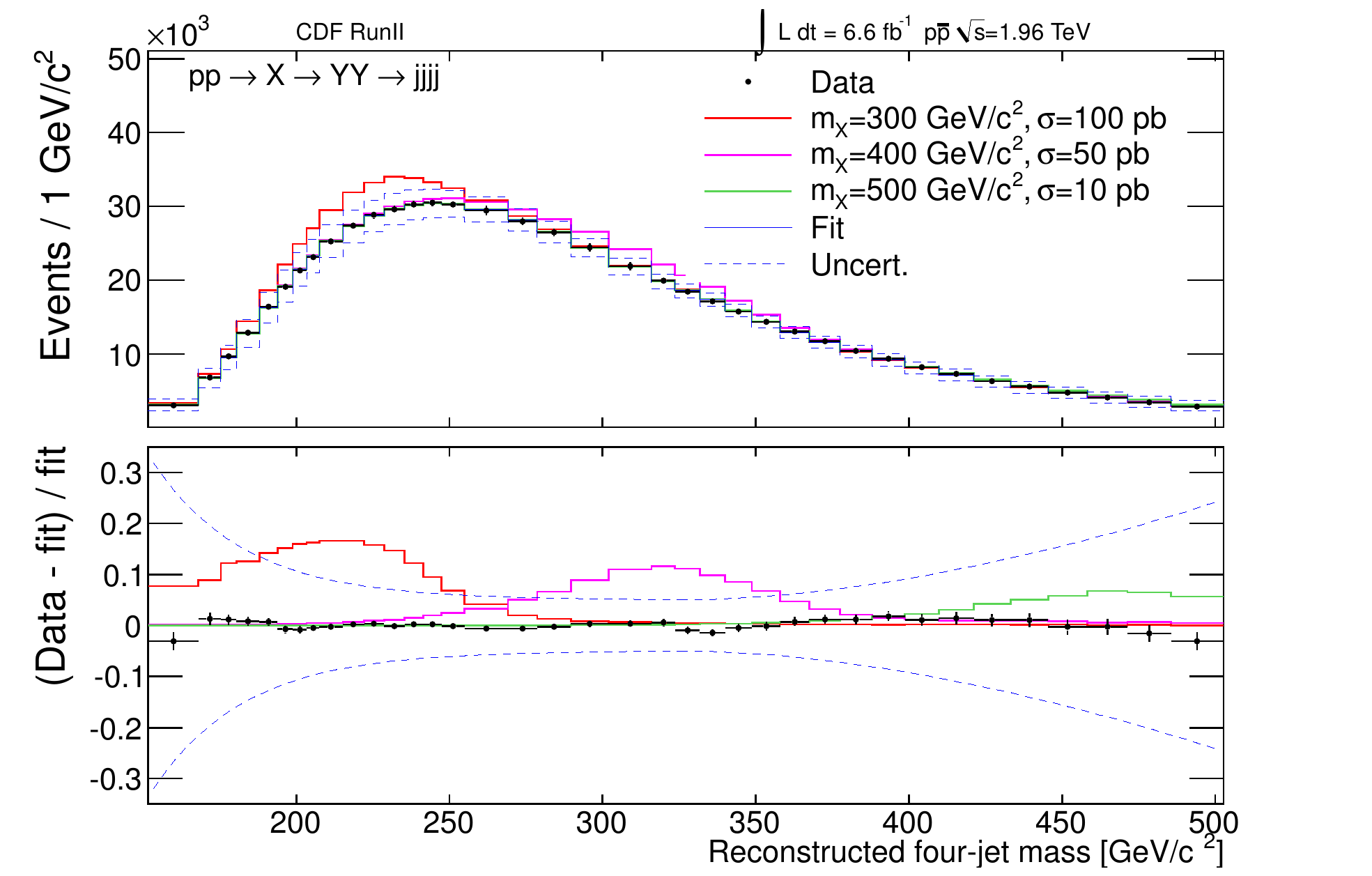}}
\put(12,50){(a)}
\put(12,26){(b)}
\end{picture}
\caption{Reconstructed  four-jet mass in events with four jets. Parametric fit and several signal
  hypotheses overlaid in (a). Relative difference
  between the observed data and the fit in each bin shown in (b).}
\label{fig:mass4j}
\end{figure}

To reconstruct the di-jet resonance, we consider the four leading jets
and evaluate the invariant mass  of each of the di-jet pairs in
the three permutations, choosing the permutation with the smallest
mass difference between the pairs.   As the pair masses are
correlated, we take the mean of the two pair masses as the estimate of
the di-jet resonance mass.  To reduce backgrounds, we require
that the relative mass difference between the two pairs is less than 50\%, and that the
production angle $\theta^*$ of the di-jet resonance in the $YY$ pair
center-of-mass frame  satisfies $\cos(\theta^*) < 0.9$. In the resonant production analysis, we calculate the four-jet
invariant mass. No specific
 $m_Y$-dependent selections are made; the requirement that the
 relative di-jet mass difference be small ensures compatibility with
 the $X\rightarrow YY$ hypothesis.  Figures~\ref{fig:mass2j} and~\ref{fig:mass4j} show the observed  di-jet and
 four-jet spectra, respectively.

% bg model

The dominant background originates from standard QCD multi-jet production.  We model this
background contribution using a parametric function which is fit to the
reconstructed mass spectrum of the observed data. The function is a piece-wise combination of a third-order
polynomial to describe the turn-on region, a third-order polynomial to
describe the peak region, and a double exponential of the form
\mbox{$f(m) = a_1 e^{-(m - a_2)^{a_{3}}/a_{4}}$} to describe the
falling spectrum. The parametric functional form was chosen to be flexible enough
to describe the multi-jet mass spectrum, but rigid enough to avoid 
accurately describing a spectrum which includes a narrow resonance, so
that in the presence of a narrow feature a signal-plus-background hypothesis would be preferred. For the di-jet mass, the
ranges used are $[35,82.5], [82.5,140]$, and $[140,700]$ GeV$/c^2$; for the
four-jet mass,  the ranges used are $[115,185], [185,330]$, and
$[330,800]$ GeV$/c^2$.  The functional form and ranges were chosen based on their
ability to accurately describe the mass spectra of simulated multi-jet
events generated by {\sc alpgen}~\cite{alpgen} version 2.10.

% systematics

The dominant source of systematic uncertainty is due to the multi-jet background
model.  The
functional form is an approximation, which even in the absence of
a narrow feature may deviate from the observed spectrum.  We estimate the
impact of these potential deviations by measuring their magnitude
in two background-enriched control samples. These
two  control samples are adjacent to the signal region and capture the
expected deviations in two independent directions.
 The first
requires a large relative di-jet mass difference, greater than 50\%,
and the second requires  $\cos(\theta^*) > 0.9$.   The observed
relative deviations are then applied to the observed spectrum in the
signal region to estimate the magnitude of spurious deviations due to
possible mismodeling.   In addition, we verify that the
fitting procedure gives an unbiased estimate of the signal rate.

An additional uncertainty is due to 
knowledge of the trigger efficiency~\cite{vhjj} extracted from the simulated signal samples,
varying from 20\% relative at $\Sigma_{\textrm{jets}} E_T=120$ GeV to 10\% above
$\Sigma_{\textrm{jets}} E_T=200$ GeV.  Uncertainties in the levels of parton radiation~\cite{rad} and in the calibration of the jet
energy and resolution modeling~\cite{jes} also contribute to
uncertainties in the trigger and selection efficiency and
reconstructed mass spectrum of the signal samples. These uncertainties are
small ($< 10\%$) relative to the fitting and trigger uncertainties.

% statistical analysis

In the non-resonant analysis, for each $Y$ mass hypothesis, we fit the
most likely value of the $Y$ pair-production
 cross section ($\sigma_{YY}$) by performing a
maximum likelihood fit of the binned di-jet mass distribution, allowing for
systematic and statistical fluctuations via template
morphing~\cite{morph}. The likelihood takes the form of 
\[L(\sigma_{YY}) = \prod_{\textrm{bin}\ i}  f^i_{\textrm{bg}}(\vec{a}) +
\sigma_{YY}\ \mathcal{L}\ \epsilon\  f_{\textrm{sig}},\]
\noindent where $f_{\textrm{bg}}(\vec{a})$ is the
parametric function with nuisance parameters $\vec{a}$ defined above to describe the background spectrum,
$f_{\textrm{sig}}$ is a normalized template of the expected shape of the signal determined
from simulated events, and $\mathcal{L} \epsilon$ is the product
of the integrated luminosity and efficiency.  No evidence is found for the presence of
pair-production of di-jet resonances and  upper limits on $Y$
pair-production at 95\% confidence level (C.L.) are set.  

Limits are calculated using the CLs~\cite{MCLimit} method by repeating the measurement on sets of simulated experiments that include signal
               contributions corresponding to various hypothetical production cross-sections, and variation of systematic uncertainties.
               The values of nuisance parameters are not fit in the experiments. The observed limits are
 consistent with expectation for the background-only hypothesis. The
 resonant analysis is very similar, but is done as a function of the $X$ mass hypothesis,
 fitting the four-jet mass distribution for the most likely value of
 $X$ production cross section, $\sigma_{X}$. 

\begin{figure}
\begin{picture}(90,40)
\put(0,0){\includegraphics[width=0.9\linewidth]{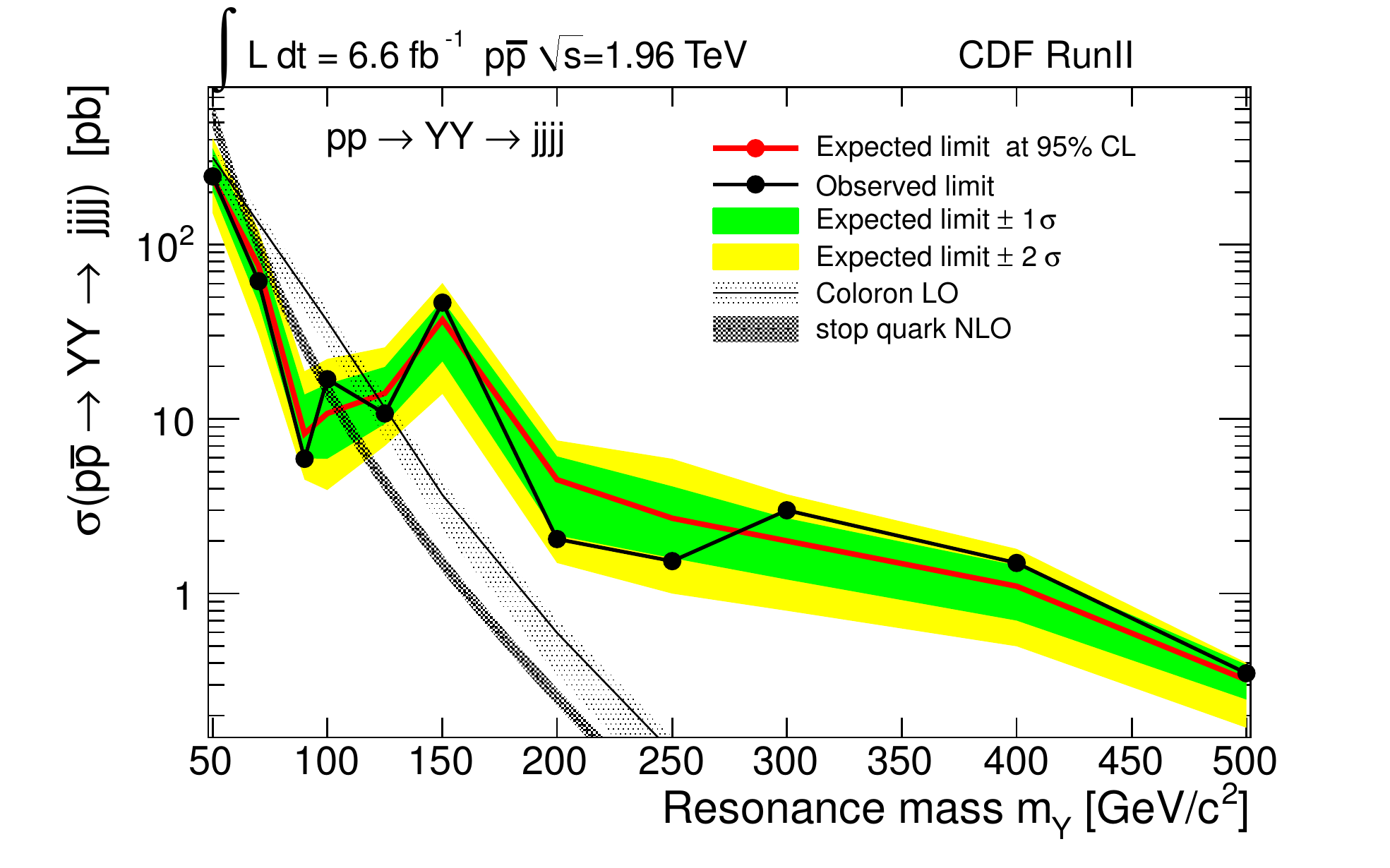}}
\put(15,12){(a)}
\end{picture}
\begin{picture}(90,63)
\put(0,0){\includegraphics[width=0.9\linewidth]{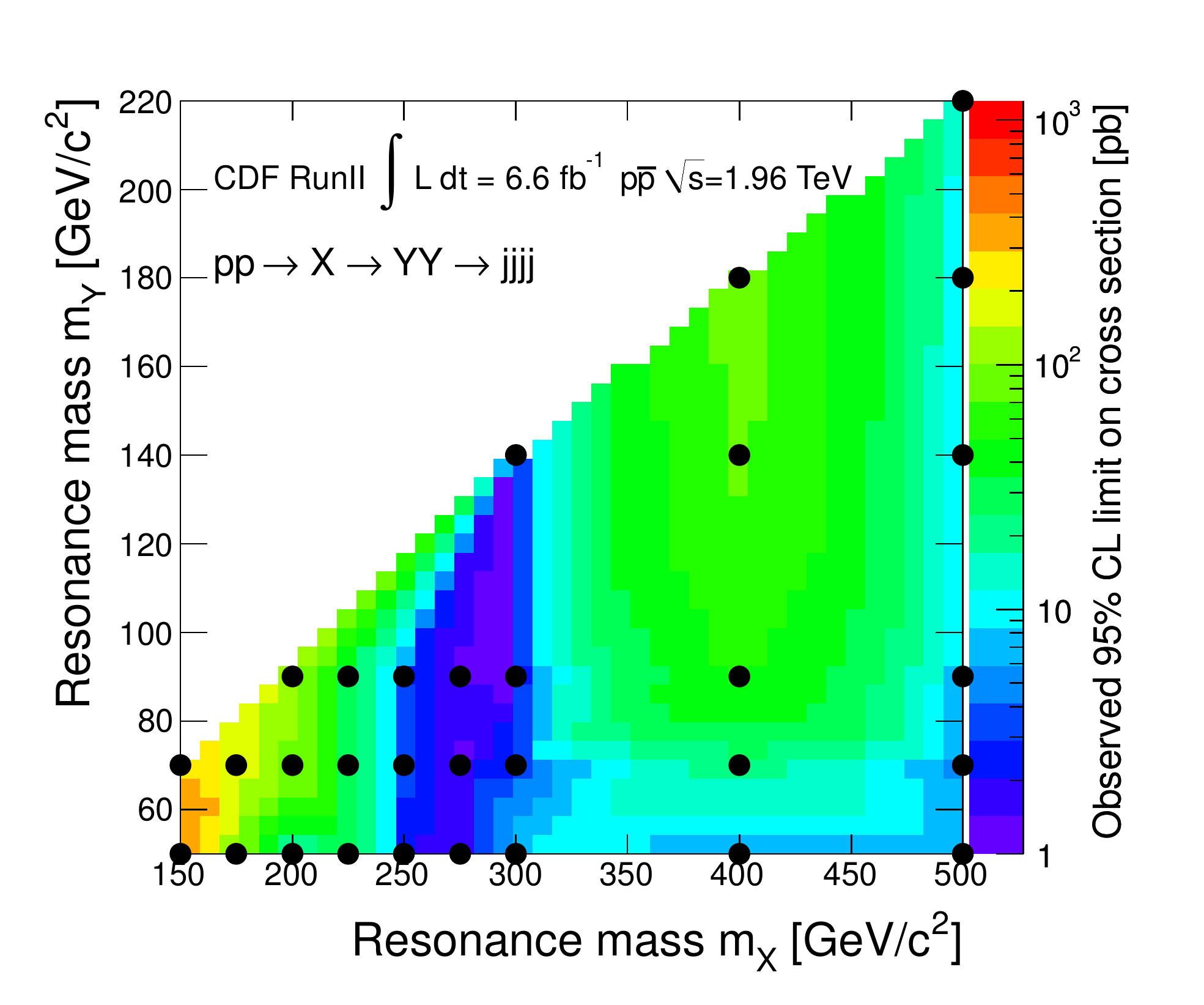}}
\put(15,27){(b)}
\end{picture}
\caption{ Upper limit on signal production rate at 95\% C.L.  
  Expected and observed upper limits on
  $\sigma(p\bar{p}\rightarrow YY \rightarrow${\it jjjj}$)$ versus $m_Y$ in
  the non-resonant analysis shown in (a). Two signal hypotheses are shown, see text
  for details. Observed limits on
  $\sigma(p\bar{p}\rightarrow X\rightarrow YY \rightarrow${\it jjjj}$)$
  versus $m_X$ and $m_Y$ shown in (b). Circles indicate the true
  values of the parameters used in each ensemble of simulated samples used to evaluate the limits; intermediate values are interpolated. }
\label{fig:lims}
\end{figure}

% interp

\begin{table}
\caption{Observed and expected 95\% C.L. upper limits on
  $\sigma(p\bar{p}\rightarrow YY\rightarrow${\it jj jj}$)$ for several values
  of $m_Y$. Also shown are theoretical predictions for coloron pair
  production ~\cite{dobrescu,kilic} or stop-quark pair production with
  $R$-parity-violating decay $\tilde{t}\rightarrow qq'$~\cite{yu}.}
\label{tab:lims}
\begin{tabular}{lcccc}
\hline\hline
Mass  & Expected & Observed & Coloron & Stop quarks\\
(GeV$/c^2$) & (pb)& (pb)& (pb)& (pb)\\
\hline
50              & 240               & 250         & 320   &  570      \\
70        & 75           & 62    & 180 & 100\\
90         & 8.2     & 5.9     & 62 & 26 \\
100        & 11    & 17     & 37 & 15 \\
125       & 14    & 11     & 11 & 4.4 \\
150       & 37   & 46     & 3.7 & 1.5 \\
200       & 4.5      & 2.0     & 0.60  & 0.25\\
250        & 2.7     & 1.5      & 0.11 & $5.4\times
10^{-2}$ \\
300        & 2.0      & 3.0   & $2.9\times
10^{-2}$& $1.3\times
10^{-2}$ \\
400        & 1.1     & 1.5     & $1.7\times10^{-3}$   & $7.2\times
10^{-4}$   \\
500       & 0.3      & 0.3      & $8.5\times19^{-5}$ & $3.6\times
10^{-5}$ \\
\hline\hline
\end{tabular}
\end{table}

\begin{table}
\caption{Observed and expected 95\% C.L. upper limits on
  $\sigma(p\bar{p}\rightarrow X\rightarrow YY\rightarrow${\it jj jj}$)$ for several values
  of $m_Y$ and $m_X$. Also shown are theoretical predictions for
  axi-gluon production assuming coupling to quarks of $C_q=0.4$~\cite{axigluon1,axigluon2}.}
\label{tab:lims4j}
\begin{tabular}{llrrr}
\hline\hline
$m_X$ & $m_Y$ & Expected & Observed & Axi-gluon\\
(GeV$/c^2$) & (GeV$/c^2$)& (pb)& (pb) & (pb)\\
\hline
\hline 150       & 50    & 641.2         & 431.1         & 5600\\ 
         & 70    & 209.6         & 270.6\\ 
\hline 175       & 50    & 66.8          & 78.9  & 3500\\ 
         & 70    & 111.5         & 163.9\\ 
\hline 200       & 50    & 13.8          & 9.5   & 2200\\ 
         & 70    & 30.4          & 91.5\\ 
         & 90    & 17.8          & 100.4\\ 
\hline 225       & 50    & 18.0          & 26.0  & 1750\\ 
         & 70    & 20.7          & 25.0\\ 
         & 90    & 20.9          & 25.3\\ 
\hline 250       & 50    & 6.2   & 2.0   & 1000\\ 
         & 70    & 4.0   & 3.6\\ 
         & 90    & 5.1   & 2.8\\ 
\hline 275       & 50    & 6.5   & 1.2   & 850\\ 
         & 70    & 7.7   & 1.3\\ 
         & 90    & 9.7   & 1.4\\ 
\hline 300       & 50    & 5.0   & 7.1   & 540\\ 
         & 70    & 2.4   & 2.6\\ 
         & 90    & 1.7   & 1.0\\ 
         & 140   & 1.8   & 1.2\\ 
\hline 400       & 50    & 15.5          & 6.8   & 170\\ 
         & 70    & 15.0          & 20.2\\ 
         & 90    & 30.6          & 52.8\\ 
         & 140   & 41.0          & 74.6\\ 
         & 180   & 46.9          & 79.1\\ 
\hline 500       & 50    & 20.7          & 6.8   & 60\\ 
         & 70    & 15.9          & 4.7\\ 
         & 90    & 17.7          & 5.9\\ 
         & 140   & 25.2          & 7.0\\ 
         & 180   & 26.7          & 8.0\\ 
         & 220   & 29.7          & 9.3\\ 
\hline\hline
\end{tabular}
\end{table}

In the non-resonant case, this analysis sets  limits on
coloron or stop-quark pair production, excluding 50-100 GeV$/c^2$ and
50-125 $/c^2$, respectively; see Table~\ref{tab:lims} and the top of Fig.~\ref{fig:lims}.  The
uncertainty on the theoretical cross-section prediction comes from two
sources summed in quadrature.
The first uncertainty is the envelope of the PDF uncertainties from
the {\sc cteq} uncertainties and an alternative PDF choice, {\sc
  mstw2008lo}~\cite{mstw} (5\% relative). The second uncertainty comes from a
variation of the  renormalization and
factorization scales by a factor of two in each direction from their
default values of the per-event mass scale. These theoretical
uncertainties are illustrated in
Figure~\ref{fig:lims}. 

In the resonant case, this analysis excludes axi-gluon ($A$)
production leading to pairs of $\sigma$ particles and four-gluon final state
for $m_A \in [150,400], m_\sigma \in [50,m_A/2]$ in the case of coupling to quarks $C_q=0.4$ (see
Table~\ref{tab:lims4j} and the bottom of Fig.~\ref{fig:lims}) which is close
to the value required to explain the top-quark $A_{\it{fb}}$
result~\cite{axigluon2}. To be consistent with this analysis, the
couplings would have to be smaller by an order of magnitude. Maintaining
consistency with the top-quark $A_{\it{fb}}$ result would require different
couplings to light quarks and heavy quarks, with the heavy-quark
coupling approaching the perturbative limit, $C_q<1$.

We thank Martin Schmaltz, Gustavo Tavares, Can Kilic, Bogdan Dobrescu,
Dirk Zerwas
and Felix Yu for useful suggestions and technical advice. 
 We thank the Fermilab staff and the technical staffs of the participating institutions for their vital contributions. This work was supported by the U.S. Department of Energy and National Science Foundation; the Italian Istituto Nazionale di Fisica Nucleare; the Ministry of Education, Culture, Sports, Science and Technology of Japan; the Natural Sciences and Engineering Research Council of Canada; the National Science Council of the Republic of China; the Swiss National Science Foundation; the A.P. Sloan Foundation; the Bundesministerium f\"ur Bildung und Forschung, Germany; the Korean World Class University Program, the National Research Foundation of Korea; the Science and Technology Facilities Council and the Royal Society, UK; the Russian Foundation for Basic Research; the Ministerio de Ciencia e Innovaci\'{o}n, and Programa Consolider-Ingenio 2010, Spain; the Slovak R\&D Agency; the Academy of Finland; and the Australian Research Council (ARC).

%\bibliography{paper}

\begin{thebibliography}{99}
\bibitem{coordinates}
CDF uses a cylindrical coordinate system with
  the $z$ axis along the proton beam axis. For a particle or a jet, pseudorapidity is $\eta
  \equiv -\ln[\tan (\theta/2)]$, where $\theta$ is the polar angle
  relative to the proton beam direction, and $\phi$ is the azimuthal
  angle  while transverse momentum $p_T = |p|\ \sin\theta$, and the
  transverse energy $E_T = E\ \sin\theta$.
\end{thebibliography}

\end{document}